%% file: ms.tex
\documentclass[]{emulateapj}
\usepackage{psfig,natbib,apjfonts}
\newcommand{\Rsun}{\mbox{\,$\rm{R_{\odot}}$}} 
\newcommand{\Lsun}{\mbox{\,$\rm{L_{\odot}}$}} 

\newcommand{\Xsun}{\mbox{\,$\rm{X_{\odot}}$}}
\newcommand{\Tstar}{\mbox{\,$T_{*}$}}
\newcommand{\Teff}{\mbox{\,$T_{eff}$}}
\newcommand{\Rstar}{\mbox{\,$R_{*}$}} 
\newcommand{\Reff}{\mbox{\,$R_{2/3}$}}

\newcommand{\Mdot}{\mbox{\,$\dot{M}$}}

\newcommand{\vinf}{\mbox{\,v$_{\infty}$}}
\newcommand{\logg}{\mbox{\,$\log{g}$}}

\newcommand{\E}[1]{\mbox{\,$\rm x 10^{#1}$}}

\newcommand{\XHe}{\mbox{\,$X_{He}$}}
\newcommand{\XC}{\mbox{\,$X_{C}$}}
\newcommand{\XN}{\mbox{\,$X_{N}$}}
\newcommand{\XO}{\mbox{\,$X_{O}$}}
\newcommand{\XNe}{\mbox{\,$X_{Ne}$}}

\newcommand{\XFe}{\mbox{\,$X_{Fe}$}}

\newcommand{\Htwo}{\mbox{\rm{H}$_2$}}
\newcommand{\HI}{\mbox{\rm{\ion{H}{1}}}}

\newcommand{\CII}{\ion{C}{2}}
\newcommand{\CIII}{\ion{C}{3}}
\newcommand{\CIV}{\ion{C}{4}}

\newcommand{\NV}{\ion{N}{5}}

\newcommand{\OV}{\ion{O}{5}}
\newcommand{\OVI}{\ion{O}{6}}
\newcommand{\NeII}{\ion{Ne}{2}}
\newcommand{\NeIII}{\ion{Ne}{3}}

\newcommand{\NeV}{\ion{Ne}{5}}
\newcommand{\NeVI}{\ion{Ne}{6}}
\newcommand{\NeVII}{\ion{Ne}{7}}
\newcommand{\NeVIII}{\ion{Ne}{8}}

\newcommand{\eg}{\emph{e.g.}}
\newcommand{\ie}{\emph{i.e.}}
\newcommand{\doublet}{$\lambda\lambda$}
\newcommand{\singlet}{$\lambda$}

\newcommand{\EBMV}{\mbox{\,$E_{\rm{B-V}}$}}

\newcommand{\kms}{\mbox{\,$\rm{km\:s^{-1}}$}}

\newcommand{\Msunyr}{\mbox{\,$\rm{M_{\odot}\:yr^{-1}}$}}

\begin{document}

\title{DISCOVERY OF NeVII IN THE WINDS OF HOT EVOLVED STARS\footnote{Based on
  observations made with the NASA-CNES-CSA Far Ultraviolet
  Spectroscopic Explorer and data from the MAST archive. FUSE is
  operated for NASA by the Johns Hopkins University under NASA
  contract NAS5-32985.}}

\author{J.E. Herald, L. Bianchi}
\vspace{1mm}
\affil{Center for Astrophysical Sciences, The Johns Hopkins University}
%\authoraddr{Bloomberg Center for Physics and Astronomy, 3400 N. Charles St., Baltimore, MD 21218-2411}
\author{D.J. Hillier}
\vspace{1mm}
\affil{Department of Physics and Astronomy, The University of Pittsburgh}
%\authoraddr{3941 O'Hara Street, Pittsburgh, PA 15260}

\begin{abstract}
We show that a strong P-Cygni feature seen in the far-UV spectra of
some very hot ($\Teff \gtrsim 85$~kK) central stars of planetary
nebulae (CSPN), which has been previously identified as \CIII\
\singlet 977, actually originates from \NeVII\ \singlet 973.  Using
stellar atmospheres models, we reproduce this feature seen in the
spectra of two [WR]-PG~1159 type CSPN, Abell 78 and NGC~2371,
and in one PG~1159 CSPN, K~1-16.  In the latter case, our analysis
suggests an enhanced neon abundance.  Strong neon features in CSPN
spectra are important because an overabundance of this element is
indicative of processed material that has been dredged up to the
surface from the inter-shell region in the
``born-again'' scenario, an explanation of hydrogen-deficient CSPN.
Our modeling indicates the \NeVII\ \singlet 973 wind feature may be
used to discern enhanced neon abundances for stars showing an
unsaturated P-Cygni profile, such as some PG~1159 stars.  We explore
the potential of this strong feature as a wind diagnostic in stellar
atmospheres analyses for evolved objects.  For the [WR]-PG~1159
objects, the line is present as a P-Cygni line for $\Teff \gtrsim
85$~kK, and becomes strong for $100 \lesssim \Teff \lesssim 155$~kK
when the neon abundance is solar, and can be significantly strong
beyond this range for higher neon abundances.  When unsaturated, \ie,
for very high \Teff\ and/or very low mass-loss rates, it is
sensitive to $\Mdot$ and very sensitive to the neon abundance. The
\NeVII\ classification is consistent with recent identification of
this line seen in absorption in many PG~1159 spectra.
\end{abstract}


\section{INTRODUCTION}\label{sec:intro}

Central stars of planetary nebulae (CSPN) represent an evolutionary
phase the majority of low/intermediate mass stars will experience.  A
small subset of CSPN have been termed ``PG~1159-[WR]'' stars, as they
represent objects transitioning from the [WR] to the PG~1159 class.
The former are objects moving along the constant-luminosity branch of
the H-R diagram which have optical spectra rich in strong emission
line features, similar to those of Wolf-Rayet (WR) stars, which represent
a late evolutionary stage of massive stars (the ``[WR]'' designation
is meant to distinguish the two).  The majority of [WR] CSPN show
prominent carbon features and are termed ``[WC]'', while a handful
show strong oxygen lines (``[WO]'').  This difference is believed to
reflect a difference in the ionization of the winds rather than in the
elemental abundances \citep{crowther:02}.  Unlike massive WR stars, CSPN
of the nitrogen-rich [WN] subtype are very rare, the two candidates
being LMC-N66 in the LMC \citep{pena:04} and PM5 in the Galaxy
\citep{morgan:03}.  The PG~1159 class marks the entry point onto the
white-dwarf cooling sequence, and these stars display mainly
absorption line profiles in the optical, as their stellar wind has
almost all but faded.  Both classes are examples of hydrogen-deficient
CSPN, which presumably make up 10-20\% of the CSPN population
(\citealp{demarco:02,koesterke:98b} and references therein), and are
believed to represent subsequent evolutionary stages based on their
similar parameters and abundances.  An explanation for the origin of
such objects is the ``born-again'' scenario (see
\citealp{iben:95,herwig:99} and references therein).  In this
scenario, helium shell flashes produce processed material between the
H- and He-burning shells (the intershell region).  This material is
enriched in He from CNO hydrogen burning, but through 3$\alpha$
process burning it also becomes enriched in C, O, Ne, and deficient
in Fe (see, \eg, \citealp{werner:04}, and references therein).  After
the star initially moves off the asymptotic giant-branch (AGB), it
experiences a late helium-shell flash, causing the star to enter a
second (or ``born-again'') AGB phase.  Flash induced mixing dredges
the processed intershell material to the surface, resulting in
H-deficient surface abundances.  When the star enters its second
post-AGB phase, its spectrum can develop strong wind features causing
it to resemble that of (massive) WR stars, perhaps because the
chemically enriched surface material increases the efficiency of
radiative momentum transfer to the wind.  Eventually, as the wind
fades, the object moves onto the white dwarf cooling sequence.  As this
happens, observable wind spectral features may only be present in the
far-UV and UV regions.

\citet{herald:04b} (hereafter, HB04) modeled the far-UV and UV spectra
of four Galactic CSPN, including Abell~78 (A78 hereafter), considered
the proto-typical transition star and candidate for the born-again
scenario.  \citet{crowther:98} classified it as a PG~1159-[WO1] star
based on its high \OVI /\CIV\ ratio.  HB04 also presented an analysis
of NGC~2371, which the authors argue is of similar nature, although
possessing a wind of even higher ionization.  In those analyses, HB04
were unable to reproduce the prominent P-Cygni feature seen in the
spectra of both stars at $\sim 975$~\AA, identified in the spectra of
A78 as \CIII\ \singlet 977 in the past \citep{koesterke:98b}.
Prominent \CIII\ \singlet 977 P-Cygni profiles do occur in the far-UV
spectra of CSPN of cooler temperatures ($\Teff
\lesssim 80$~kK, see, \eg, \citealp{herald:04a}), as well as in
massive hot stars.  However, effective temperatures of the transition
objects (such as A78) are found to be very high (\ie, $\gtrsim 90$~kK)
from both stellar atmospheres codes (\eg, \citealp{werner:03}, HB04)
and nebular line analyses (\eg, \citealp{kaler:93,grewing:90}).  HB04
noted that although this feature has been assumed to be from \CIII, a
strong transition from an ion of relatively low ionization potential
(48 eV) in such highly-ionized winds was strongly questionable.  HB04
investigated and excluded the possibility that the line originated
from a highly ionized iron species.

\citet{werner:04} reported the identification of a narrow absorption
feature at \singlet 973 in the spectra of several PG~1159 stars as
\NeVII.  In some cases, this line is superimposed on a broad P-Cygni
feature, which they identified as \CIII\ \singlet 977.  Given that the
high ionization potential of \NeVII\ (207 eV) is more consistent with
the conditions expected in objects of such high temperatures, we were
motivated to include neon in the model atmospheres of A78 and NGC~2371
presented in HB04.  Neon had not been included in any previous
modeling.  We present the results of this analysis, and report that
neon can adequately account for this hitherto unexplained wind
feature.  Additionally, we show that \NeVII\ is also responsible for
the broad P-Cygni feature seen in the spectra of the PG~1159 star
K~1-16.  We also investigate, with a grid of models, the usefulness of
this line as a wind diagnostic for very hot CSPN.  As the neon
abundance is also of interest with respect to massive stars with
winds, this work may have application to the study of the hottest
Wolf-Rayet stars as well.  This paper is arranged as follows: the
observations are described in \S~\ref{sec:obs}.  The models are
described in \S~\ref{sec:modeling}.  Our results are discussed in
\S~\ref{sec:discussion} and our conclusions in
\S~\ref{sec:conclusions}.

\section{OBSERVATIONS AND REDUCTION}\label{sec:obs}

The data sets utilized in this paper are summarized in
Table~\ref{tab:obs}. For NGC~2371 and K~1-16, we have used far-UV data
from \emph{Far-Ultraviolet Spectroscopic Explorer} (FUSE), and for
A78, from the \emph{Berkeley Extreme and Far-UV Spectrometer} (BEFS).
For NGC~2371, we have also made use of a UV \emph{International
Ultraviolet Explorer} (IUE) spectrum.  The data characteristics, and
the reduction of the FUSE data, are described in HB04.  The data were
acquired from the MAST archive.

For K~1-16, the FUSE data were reduced in a similar manner as described in
HB04, except with the latest version of the FUSE pipeline (CALFUSE
v2.4).  The count-rate plots show that the star was apparently out of
the aperture during part of the observation, and data taken during
that period was omitted from the reduction process.

The radial velocities of NGC~2371 and A78 are $+20.6$ and $+17$~\kms,
respectively \citep{acker:92}.  All observed spectra presented in this
paper have been velocity-shifted to the rest-frame of the star based
on these values.

The far-UV spectra of our sample are shown in Fig.~\ref{fig:fuv_all},
along with our models (described in \S~\ref{sec:modeling}).  They are
mainly dominated by two strong P-Cygni features - \OVI\ \doublet
1032,38 and \NeVII\ \singlet 973.  Both features are saturated in the
spectra of NGC~2371 and A78, while they are unsaturated in that of
K~1-16.  The numerous absorption lines seen are due to the Lyman and
Werner bands of molecular hydrogen (\Htwo), which resides in both the
interstellar and circumstellar medium (discussed in HB04).

Close inspection of the \NeVII\ P-Cygni profile reveals that there
does appear to be some absorption due to \CIII\ \singlet 977 in each
case, as well as emission in the case of NGC~2371.  This apparently
arises from absorption by cooler carbon material in the circumstellar
environment, perhaps similar to the ``carbon curtain''
\citet{bianchi:87b} invoked to explain the similar \CII\ \doublet
1334.5,1335.7 features seen in the spectra of NGC~40 (that CSPN has a
temperature of $\Teff = 90$~kK, as estimated from the UV spectrum, too
hot for \CII\ to be present in the stellar atmosphere).

\section{MODELING}\label{sec:modeling}

To analyze the spectra of our sample, we have computed non-LTE
line-blanketed models which solve the radiative transfer equation
in an extended, spherically-symmetric expanding atmosphere.
The models are identical to those described in HB04, except neon
is now included in the model atmospheres.  The reader is
referred to that work for a more detailed description of the models,
here we give only a summary.

The intense radiation fields and low wind densities of CSPN invalidate
the assumptions of local thermodynamic equilibrium, and their extended
atmospheres necessitate a spherical geometry for solving the radiative
transfer equation.  To
model these winds, we have used the CMFGEN code
\citep{hillier:98,hillier:99b,hillier:03}. The detailed workings of
the code are explained in the references therein.  To summarize, the
code solves for the non-LTE populations in the co-moving frame of
reference.  The fundamental photospheric/wind parameters include
\Teff, \Rstar, \Mdot, the elemental abundances and the velocity law
(including \vinf).  The \emph{stellar radius} (\Rstar) is taken to be
the inner boundary of the model atmosphere (corresponding to a
Rosseland optical depth of $\sim20$).  The temperature at different
depths is determined by the \emph{stellar temperature} \Tstar, related
to the luminosity and radius by $L = 4\pi\Rstar^2\sigma\Tstar^4$,
whereas the \emph{effective temperature} (\Teff) is similarly defined
but at a radius corresponding to a Rosseland optical depth of 2/3.
The luminosity is conserved at all depths, so $L =
4\pi\Reff^2\sigma\Teff^4$.  We assume what is essentially a standard
velocity law $v(r) = \vinf(1-r_0/r)^\beta$ where $r_0$ is roughly
equal to \Rstar, and $\beta =1$.

For the model ions, CMFGEN utilizes the concept of ``superlevels'',
whereby levels of similar energies are grouped together and treated as
a single level in the rate equations \citep{hillier:98}.  Ions and the
number of levels and superlevels included in the model calculations
are listed in Table~\ref{tab:ion_tab}.  The atomic data references are
given in HB04, except for neon (discussed in \S~\ref{sec:neon}).  The
parameters of the models presented here 
are given in Table~\ref{tab:mod_param_dist}.

\subsection{Abundances}\label{sec:abund}

Throughout this work, the nomenclature $X_i$ represents the mass
fraction of element $i$, ``\Xsun'' denotes the solar abundance, with
the values for ``solar'' taken from \citet{grevesse:98} (their
solar abundance of neon is 1.74\E{-3} by mass).  As explained in
HB04, an abundance pattern of \XHe,\XC,\XO = 0.54, 0.36, 0.08 was
adopted to model these hydrogen deficient objects.  The nitrogen
abundance was taken to be \XN = 0.01, and solar values were adopted
for the other elements, except for iron.  HB04 and \citet{werner:03}
found a sub-solar iron abundance was required to match observations of
A78, and our models of that star have $\XFe = 0.03\Xsun$.

\subsection{Neon}\label{sec:neon}

The prominent far-UV \NeVII\ feature arises from the $2p^1P^o - 2p^2
\:^1D$ transition.  As discussed by \citet{werner:04}, there is some
uncertainty in the corresponding wavelength. We adopt 973.33 \AA, the
value found in the Chianti database \citep{young:03} and which was was
measured by \citet{lindeberg:72}.  The corresponding lower and upper
level energies are 214952.0 and 317692.0~$\rm{cm^{-1}}$, respectively.

The neon atomic data was primarily taken from the Opacity Project
\citep{seaton:87,opacity:95,opacity:97} and the Atomic Spectra
Database at the NIST Physical Laboratory. For \NeVII, energy level data
have been taken from NIST with the exception of the $2p^2 \:^1D$
level, for which we have used the value from the Chianti database.
Individual sources of atomic data (photo-ionization and cross-sections)
include the following: \citet{luo:89a} (\NeV), \citet{tully:90} (\NeVII),
and \citet{peach:88} (\NeVIII).

\section{RESULTS}\label{sec:discussion}

The goal of this work was to test whether the inclusion of neon in the
models of HB04 could account for the strong P-Cygni feature appearing
at $\sim975$~\AA\ in the spectra of two transition stars (NGC~2371 and
A78), which previously has lacked a plausible explanation (HB04).  The
previous common identification with \CIII\ \singlet 977 was questioned
by HB04, as the presence of this ion would imply a much lower \Teff,
inconsistent with other spectral diagnostics.  \citet{koesterke:98b}
speculated that for A78, neglected iron lines might sufficiently
cool the outer layers of the (otherwise hot) wind to allow for the
formation of \CIII.  However, HB04 computed models which included highly
ionized iron, and excluded this explanation for the observed feature.
As we discuss below, we find that this P-Cygni line originates from \NeVII\
in both stars.  We also show that this is the case for a PG~1159 star,
K~1-16.  Additionally, we explore the usefulness of this feature as a
diagnostic for stellar parameters, which is very important given the
scarcity of diagnostic lines at high effective temperatures (discussed
by HB04).

\subsection{NGC~2371 \& A78}\label{sec:transition}
We initially re-calculated the NGC~2371 and A78 best-fit models of
HB04 (Table~\ref{tab:mod_param_dist}) including neon at solar
abundance.  The resulting models reproduced the \singlet 973 P-Cygni
feature at a strength comparable to the observations, (although a bit
weak in both cases), showing that \NeVII\ is indeed responsible for
this line.  We also computed models with higher neon abundances, which
is a predicted consequence of the ``born-again'' scenario (see
\S~\ref{sec:diag}).  In Fig.~\ref{fig:fuv_all}, we show the $\XNe =
10\Xsun$ models, which are nearly indistinguishable from the solar
abundance models (see \S~\ref{sec:diag}).  We have applied the effects
of \HI\ and \Htwo\ absorption to the model spectra as described in
HB04.  The feature is weaker in the observations of A78 than in those
of NGC~2371, and its blue P-Cygni edge is more severely affected by
absorption from \Htwo, making assessments of the quality of its model
fit more uncertain.  We note here that the parameters derived by HB04
were determined from a variety of diagnostic lines, and the listed
uncertainties take into account all the different adjustments needed
to fit them all, not just the ones shown here.  In addition, the
inclusion of neon in the calculation does not change the ionization
significantly for other abundant ions in the wind (as can be seen in
Fig.~\ref{fig:fuv_all}, where the HB04 models are also plotted), thus
there was no need for a revision of the stellar parameters.

\subsection{K~1-16}\label{sec:k116}

Based on the similar parameters (\Teff\ and abundances) of NGC~2371
and A78 to those of PG~1159 stars discussed by \citet{werner:04}, we suspected
that the broad P-Cygni profile identified therein as \CIII\ \singlet
977 in the spectra of a few of their objects originated from \NeVII\
as well.  \citet{werner:04} do classify a \NeVII\ line from their
(static) models, but only attribute this identification to a narrow
absorption feature, while identifying the broad P-Cygni
wind feature as \CIII.

\begin{figure}[htbp]
\begin{center}
\epsscale{.8}
%\begin{turn}{90}
\rotatebox{0}{
\plotone{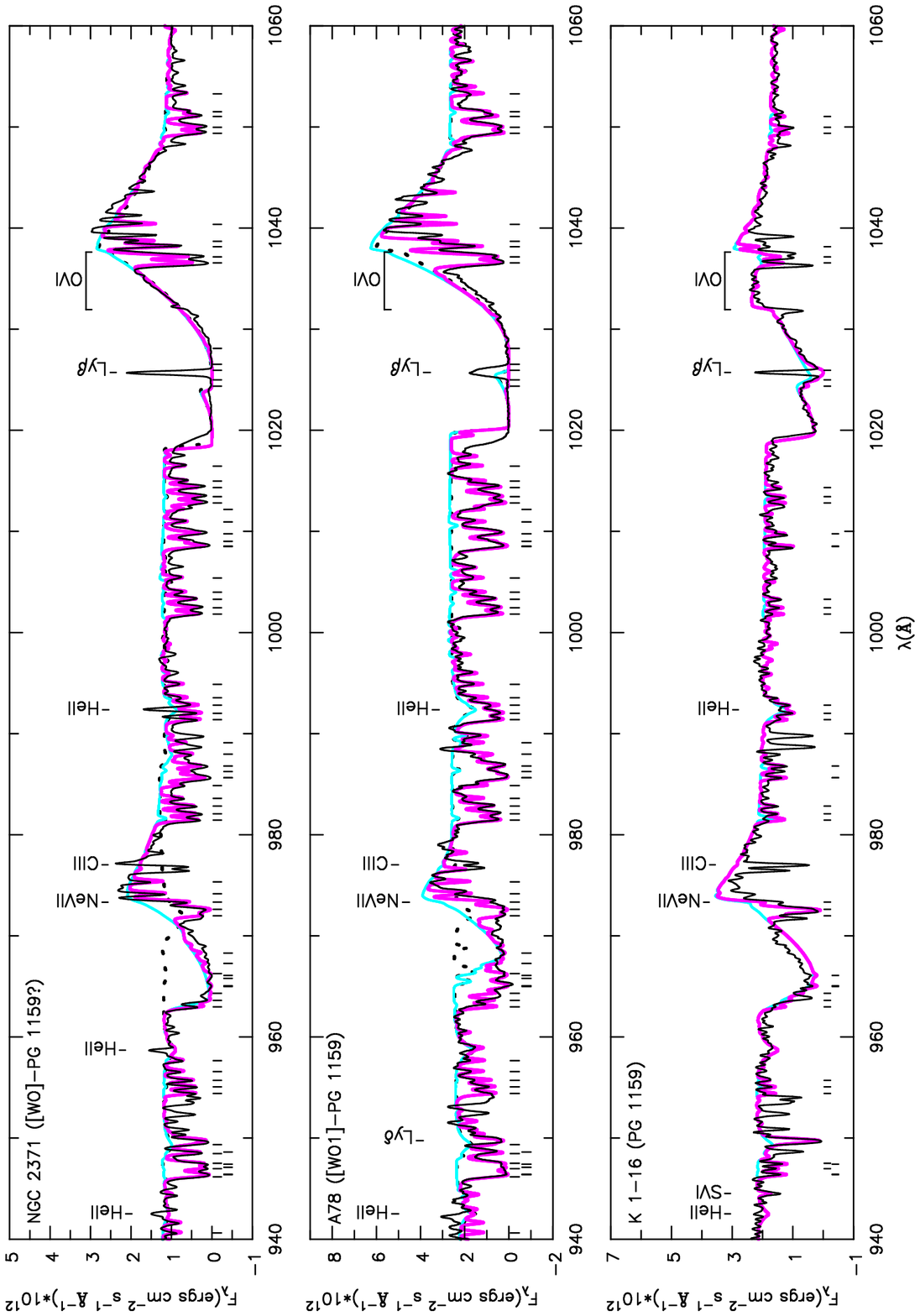}}
%\end{turn}
\caption{The FUSE observations (black) along with our stellar models
 (with $\XNe=10\Xsun$), both with (pink/dark gray) and without (aqua/light
 gray) the effects of the hydrogen absorption models applied (see
 text).  The models indicate \NeVII\ to be responsible for the P-Cygni
 feature seen at $\sim 975$~\AA.  We also show the original HB04
 models of NGC~2371 and A78 (black dotted), which did not include neon
 in the model atmospheres.  }\label{fig:fuv_all}
\end{center}
\end{figure}

\begin{figure}[htbp]
\begin{center}
\epsscale{0.4}
%\begin{turn}{90}
\rotatebox{90}{
\plotone{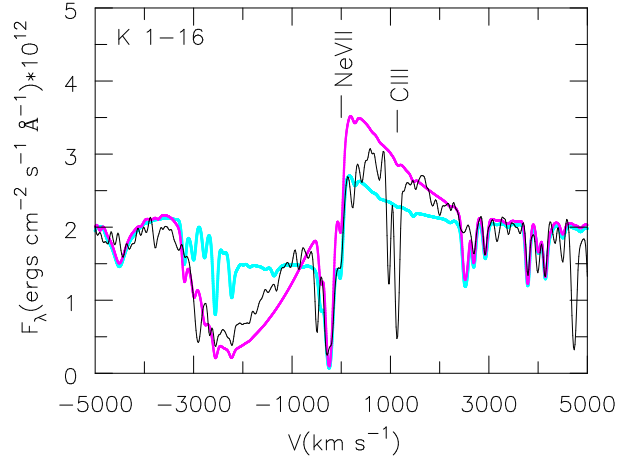}}
%\end{turn}
\caption{FUSE spectrum of the \NeVII\ \singlet 973 profile (in
  velocity space) of K~1-16 (black) along with two model spectra, one
  with $\XNe = 1\Xsun$ (solar), and one with $\XNe = 10\Xsun$
  (aqua/light gray and pink/dark gray, respectivly).  Effects of
  interstellar \HI\ and \Htwo\ absorption have been applied to the
  model spectra (\S~\ref{sec:k116}).  The unsaturated profile is very
  sensitive to the neon abundance, and indicates a super-solar
  abundance in this object.  }\label{fig:k116_XNe}
\end{center}
\end{figure}

We decided to test this hypothesis for the case of the PG~1159 star
K~1-16. \citet{koesterke:98b} determined the following parameters for
K~1-16 from a hydrostatic analysis: $\Tstar = 140$~kK, $\log{L/\Lsun}
= 3.6$ (which imply $\Rstar = 0.11$~\Rsun), $\logg=6.1$, \XHe, \XC,
\XO = 0.38, 0.56, 0.06, and the following parameters from a wind-line
analysis: $\vinf = 4000$~\kms\ and $\log{\Mdot} = -8.1$~\Msunyr\ from
the \OVI\ resonance lines.  Although the mass-loss rate is lower, the
other parameters are close to those NGC~2371, so we first took the
parameters of our NGC~2371 model shown in Fig.~\ref{fig:fuv_all}, and
scaled them to K~1-16's radius of $\Rstar = 0.11$~\Rsun\ as determined
by \citet{koesterke:98b}.  Further scaling of the model flux is needed
to match the observed flux levels of K~1-16, and this scaling is
equivalent to the star lying at a distance of 2.05~kpc.  The only
distance estimates to this star are statistical (based on nebular
relations), ranging from 1.0 to 2.5~kpc (\citealp{cahn:92,maciel:84},
respectively).  The FUSE spectrum of K~1-16 (Fig.~\ref{fig:fuv_all})
shows a unsaturated \OVI\ P-Cygni profile of comparable strength to
the \NeVII\ feature (which is also unsaturated), in contrast to that
of NGC~2371, where the profiles are saturated and that of \OVI\ is
stronger than that of \NeVII.  We therefore decreased the mass-loss
rate of the scaled model with solar neon abundance until the \OVI\
feature was fit adequately (the resulting parameters of the best-fit
model for K~1-16 are listed in Table~\ref{tab:mod_param_dist}).
The resulting model's \NeVII\ feature is
unsaturated and is weak compared to the observations, as shown in
Fig.~\ref{fig:k116_XNe}.  We also computed an enriched model with
$\XNe =10$~\Xsun\ (also shown).  As the figure illustrates, the
feature (unsaturated in this case) is now very sensitive to the
abundance (unlike the cases of NGC~2371 and A78), and the profile of
the Ne-enriched model is now too strong.  Fig.~\ref{fig:fuv_all} shows
both the \NeVII\ and \OVI\ lines of the Ne-enriched K~1-16 model.

Our models undoubtedly show that \NeVII, not \CIII, accounts for the
broad P-Cygni profile seen in the PG~1159 stars as well.  Furthermore,
our models demonstrate how this feature can be used to detect a
supersolar neon abundance in the case of an unsaturated \NeVII\
\singlet 973 profile.  Although these results suggest a supersolar
abundance for K~1-16, a more complete photospheric/wind-line analysis
should be performed, given that other parameters influence the strength
of this line (\S~\ref{sec:diag}).

To account for the interstellar \HI\ absorption, we have used the \HI\
parameters utilized by \citet{kruk:98} ($\log{N(\HI)} =
20.48$~cm$^{-2}$, $b=20$~\kms).  We modeled the \Htwo\ absorption
using $\log{N(\Htwo)} = 16.0$~cm$^{-2}$, which produces fits adequate
for our purpose (Figs.~\ref{fig:fuv_all} and \ref{fig:k116_XNe}). We
have assumed \HI\ and \Htwo\ gas temperatures of 80~K, and used the
same methods described in HB04 to calculate the absorption profiles.
We found a slightly higher reddening value than that of
\citet{kruk:98} to produce better results ($\EBMV = 0.025$
vs. 0.02~mag).  We note that the far-UV spectrum of K~1-16 shows very
little absorption from \Htwo\ compared to other CSPN (\eg,
\citealp{herald:02,herald:04a,herald:04b}), presenting a very
``clean'' example of a far-UV CSPN spectrum.

\subsection{\NeVII\ \singlet 973 as a DIAGNOSTIC}\label{sec:diag}

To investigate the potential of this feature as a diagnostic of
stellar parameters, we computed exploratory models varying either the
neon abundance, the mass-loss rate or the temperature of the A78 and
NGC~2371 models (while keeping the other parameters the same) to study
the sensitivity of this line to each parameter.

\begin{figure}[htbp]
\begin{center}
\epsscale{0.5}
%\begin{turn}{90}
\rotatebox{0}{
\plotone{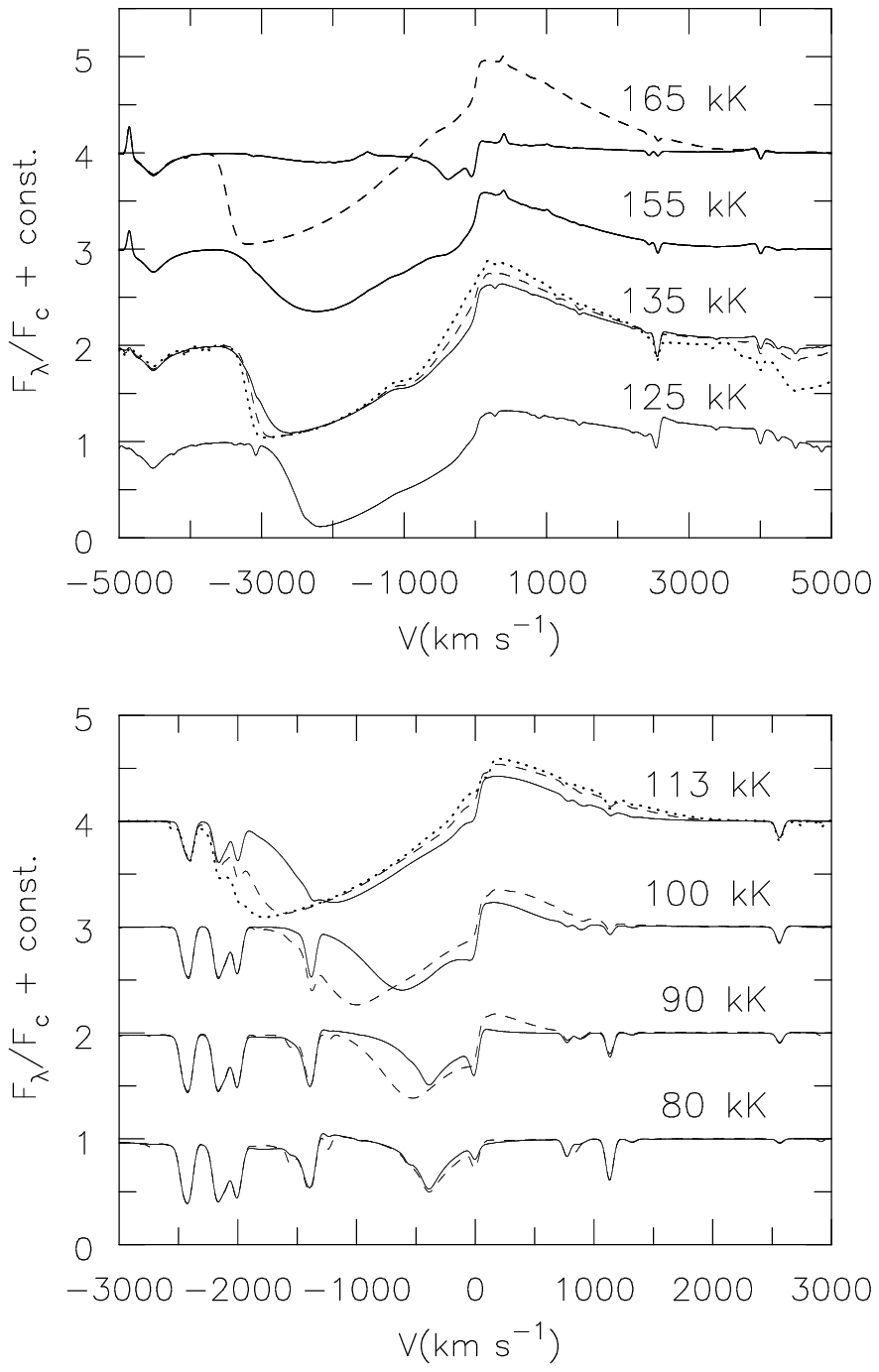}}
%\end{turn}
\caption{The effects of varying \Teff\ and \XNe\ of our NGC~2371 model
  (top) and A78 model (bottom) on \NeVII\ \singlet 973.  The solid,
  dashed, and dotted lines represent models with neon abundances of
  $1\Xsun$ (solar), $10\Xsun$, and $50\Xsun$, respectively.  When
  saturated (\eg, in the $\Teff=135$~kK model), the feature is not
  very sensitive to $\XNe$.  It is sensitive to $\Teff$, and becomes
  pure absorption for $\Teff \lesssim 85$~kK and is of significant
  strength for $100 \lesssim \Teff \lesssim 160$~kK.  The feature is
  weak in the solar neon abundance $\Teff=165$ model, but strengthens
  dramatically when $\XNe = 10\Xsun$, illustrating the abundance sensitivity of
  the line in some parameter regimes where it is
  unsaturated.  The spectra are normalized (and in velocity space),
  and convolved with a 0.2~\AA\ Gaussian for clarity.
  }\label{fig:diag}
\end{center}
\end{figure}

\begin{figure}[htbp]
\begin{center}
\epsscale{0.4}
%\begin{turn}{90}
\rotatebox{90}{
\plotone{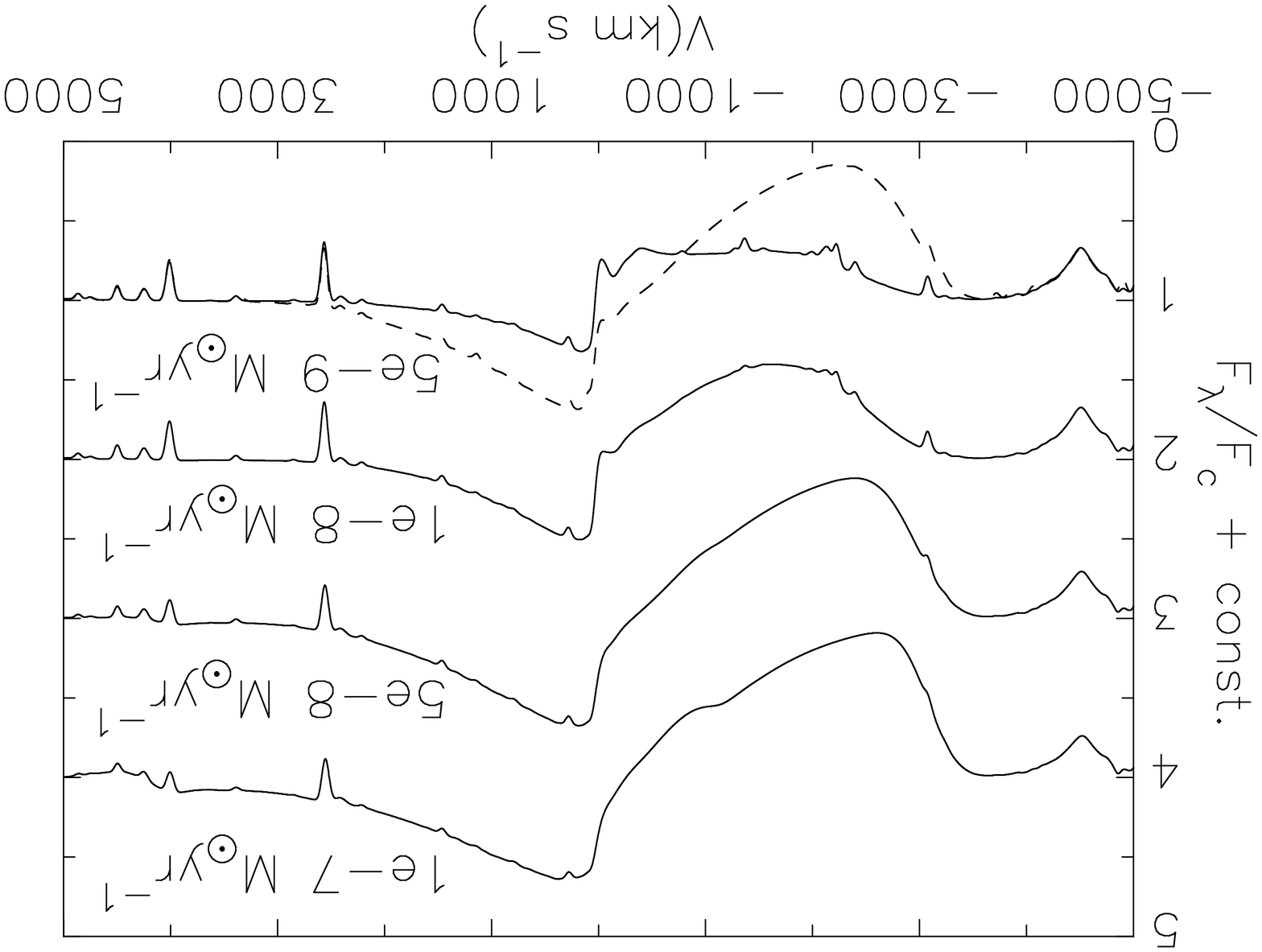}}
%\end{turn}
\caption{The effects of varying \Mdot\ and \XNe\ of our K~1-16 model ($\Teff =
  135$~kK, $\Rstar = 0.11$~\Rsun) on \NeVII\ \singlet 973.  The
  solid and dashed lines represent models with neon abundances of
  $1\Xsun$ (solar) and $10\Xsun$, respectively.  The feature is very
  sensitive to \Mdot\ when not saturated (\eg, the
  $\Mdot=5\E{-9}$~\Msunyr\ model). The spectra are normalized, and
  convolved with a 0.2~\AA\ Gaussian for clarity.
  }\label{fig:k116_mdot}
\end{center}
\end{figure}

Evolutionary calculations of stars experiencing the ``born-again''
scenario (\eg, see \citealp{herwig:01}) predict a neon abundance of
$\sim2$\% by mass in the intershell region, produced via
$\:^{14}\rm{N}(\alpha,\gamma)\:^{18}\rm{F}(\rm{e}^{+}\nu)\:^{18}\rm{O}(\alpha,\gamma)\:^{22}\rm{Ne}$.
This material later gets ``dredged up'' to the surface, resulting in a
surface abundance enhancement of up to 20 times the solar value.  We
have thus calculated models with super-solar neon abundances (with
$\XNe = 10$~\Xsun, and 50~\Xsun) to gauge the effects on
the \singlet 973 feature (shown in Fig.~\ref{fig:diag}).  As expected,
because the feature is nearly saturated in both cases, the line shows
only a weak dependence on \XNe.  The enriched models do result in a
better fit than the solar abundance models.  However, given the
sensitivity of this line to $\Teff$ (discussed below) and the
uncertainty in this parameter (see HB04), we cannot make a definitive
statement about the neon abundance of these transition objects based
solely on this wind line.  On the other hand, the strength of the
\NeVII\ line depends dramatically on the neon abundance in parameter
regimes where it is not saturated, \eg, for very high \Teff\ or very
low mass-loss rates, as discussed below and shown in
Figs.~\ref{fig:k116_XNe}-\ref{fig:k116_mdot}. 

To test the sensitivity of the \singlet 973 feature to \Mdot, we have
computed a range of models varying the mass-loss rates of our models
(with solar neon abundance) while keeping the other parameters fixed
for each.  We find virtually no change while the \singlet 973 profile
remains saturated, until the change in \Mdot\ induces a significant
change in the ionization of the wind.  This is shown for the model parameters
of K~1-16 in Fig.~\ref{fig:k116_mdot},
where the \NeVII\ profile is essentially unchanged for $5\E{-8} <
\Mdot < 1\E{-7}$~\Msunyr, and then weakens dramatically as \Mdot\ lowered
to $1\E{-8}$~\Msunyr.  However, if the atmosphere is
Ne-enriched, this limit could be significantly lower, as illustrated
in the $5\E{-9}$~\Msunyr\ ($\Teff=135$~kK) case.

\begin{figure}[htbp]
\begin{center}
\epsscale{0.9}
%\begin{turn}{90}
\rotatebox{0}{
\plotone{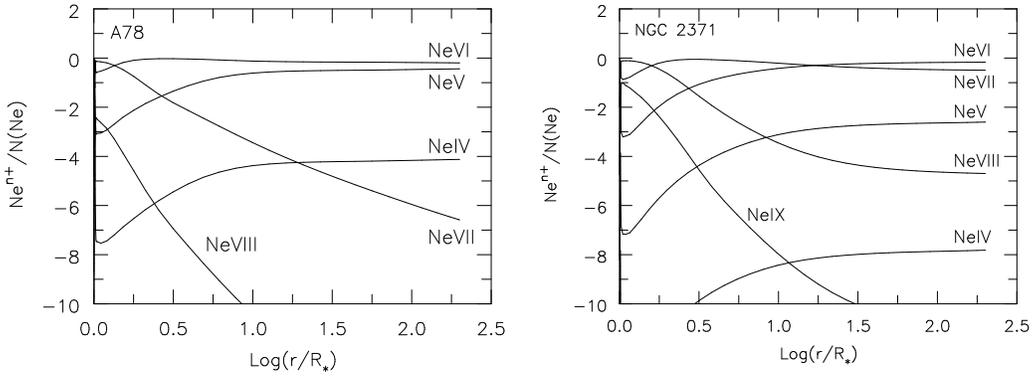}}
%\end{turn}
\caption{The neon ionization structures of our models for A78 ($\Teff
  \simeq 115$~kK) and NGC~2371 ($\Teff \simeq 135$~kK)
  as a function of distance from the stellar surface.  For the cooler
  object, \NeVII\ is only dominant deep in the wind, while in the
  hotter model it is dominant further out.  
}\label{fig:ion}
\end{center}
\end{figure}

The ionization structures of neon for the A78 and NGC~2371 models
(with $\XNe=1\Xsun$) are shown in Fig.~\ref{fig:ion}.  In the cooler
A78 model, \NeVII\ is only dominant deep in the wind, with \NeVI\
being dominant in the outer layers.  We have explored the temperature
sensitivity of the feature by adjusting the luminosity of our default
models while keeping the other parameters (\ie, \Rstar\ and \Mdot) the
same.  The results are shown in Fig.~\ref{fig:diag}.  For the A78
model parameters, the \NeVII\ \singlet 973 wind feature weakens as the
temperature is decreased, becoming insignificant for $\Teff \lesssim
85$~kK.  For the NGC~2371 model parameters, it weakens significantly
as the temperature is lowered from $\simeq 130$~kK to $\simeq 110$~kK.
The profile is fairly constant for $130 \lesssim \Teff \lesssim
145$~kK, and then starts to weaken as the temperature is increased, as
\NeVII\ ceases to be dominant in the outer wind (around $\Teff \simeq
150$~kK), and becomes a pure absorption line for $\Teff \gtrsim
170$~kK.  However, these thresholds are dependent on the neon
abundance, as illustrated by the $\Teff = 165$~kK models.  For that
temperature, the solar neon abundance model shows only a weak,
unsaturated P-Cygni profile, but in the $\XNe = 10\Xsun$ model, the
line increases dramatically, becoming saturated again.  This shows the
potential of this line to exist in strength over a wide range of
effective temperatures, as well as being an Ne-abundance diagnostic.
The presence of \NeVII\ as a P-Cygni profile sets a lower limit to
\Teff\ ($\sim 80$~kK), quite independent of the neon abundance.

\begin{figure}[htbp]
\begin{center}
\epsscale{0.9}
%\begin{turn}{90}
\rotatebox{0}{
\plotone{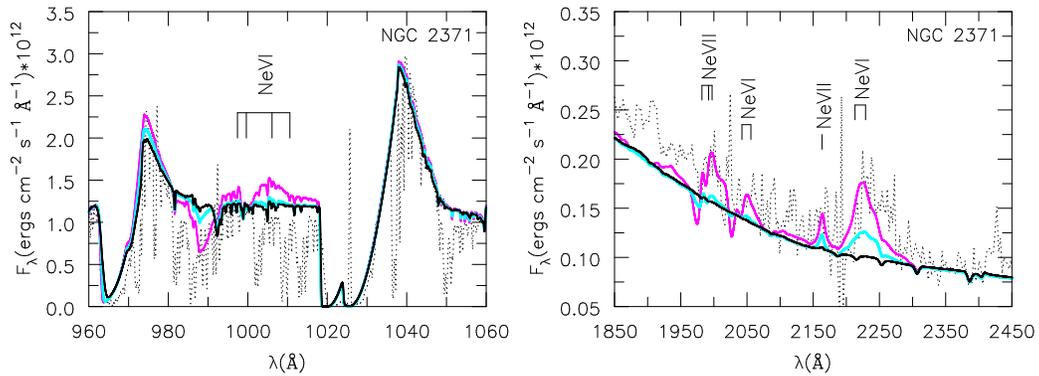}}
%\end{turn}
\caption{Additional neon diagnostics: The NGC~2371 FUSE (left) and IUE
  (right) observations
  are shown (dotted) along with models with
  $\XNe = 1\Xsun$ (black), $\XNe = 10\Xsun$ (aqua/light gray) and $\XNe
  = 50\Xsun$ (pink/dark gray). The labeled features are not apparent
  in the solar abundance model, but become significant as the neon
  abundance is increased.  }\label{fig:neon_lines}
\end{center}
\end{figure}

Although our modeling indicates the \NeVII\ \singlet 973 feature may
not be useful in diagnosing a super-solar neon abundance in the case
of NGC~2371 (because it is saturated) and A78 (because its blue edge
is obscured by \Htwo\ absorption), it does show other neon
transitions in the far-UV and UV which do not produce significant
spectral features for a solar neon abundance, but do for enriched neon
abundances (see examples in Fig.~\ref{fig:neon_lines}).  The strongest
examples are the \singlet 2213.13 and \singlet 2229.05 transitions
from the $3d \:^2D - 3p \:^2P^o$ \NeVI\ triplet which become evident
in models of both objects when $\XNe = 10\Xsun$.  For this abundance,
\NeVI\ $3p \:^2P^o - 3s \:^2S$ transitions are also seen at \singlet
2042.38 and \singlet 2055.94 in the A78 models, and the \NeVII\ $3s^1S
- 3p^1 P^o$ transition (\singlet 3643.6) in the models of NGC~2371
(this line has been used by \citealp{werner:94} to deduce
enhanced neon abundances in a few PG~1159 stars, including K~1-16).
Also seen in the $\XNe = 10\Xsun$ models of NGC~2371 is the \NeVII\ $3p^3P^o -
3d \:^3D$ multiplet, which \citet{werner:04} observed at positions
shifted about 6~\AA\ blueward of the wavelengths listed in the NIST
database (the strongest observed component occurs at \singlet 3894).
For $\XNe = 50\Xsun$, in the model of NGC~2371, \NeVI\ $2p^2 \:^4P - 2p^2
\:^2P^o$ features are seen from 993~\AA\ to 1011~\AA, as well as blend
of \NeV - \NeVII\ transitions at $\sim2300$~\AA.  Although the
resolution and/or quality of the available IUE data in this range are
not sufficient to rigorously analyze these lines, we note that the
observations seem to favor a higher neon abundance, based on a
significantly strong feature at 2230~\AA\ in the IUE observations of
NGC~2371 that is only matched by the $\XNe = 50\Xsun$ model.

We note that introducing neon at solar abundance in the model
atmospheres of these objects does not have a significant impact on the
ionization structure of other relevant ions in the wind.  For A78, at
$\XNe = 10\Xsun$, the ionization structures of other elements changes
slightly, but not enough to result in spectral differences.  Very high
neon abundances ($\XNe = 50\Xsun$) result in a less ionized wind,
with, for example, the \OV\ \singlet 1371 feature strengthening as
this ion becomes more dominant in the outer parts of the wind.  The
neon ionization structure is most dramatically affected - with \NeV\
becoming the dominant ion in the outer parts of the wind.  For
NGC~2371, introducing neon at solar abundance reduces the ionization
of the wind slightly, with the effect becoming more significant at
$\XNe \ge 10\Xsun$ when it leads to stronger \NV\ and \OV\ features at
UV wavelengths.  Thus, high neon abundances can significantly
influence the atmospheric structure, and fitting a spectrum using the
same set of non-neon diagnostics with a model with an enriched neon
abundance generally seems to require a higher luminosity.

\section{CONCLUSIONS}\label{sec:conclusions}

We have shown that the strong P-Cygni wind feature seen around
975~\AA\ (hitherto unidentified or mistakenly identified as \CIII\
\singlet 977) in the far-UV spectra of very hot ($\Teff \gtrsim
100$~kK) CSPN can be reproduced by models which include neon in the
stellar atmosphere calculations.  We have demonstrated this
identification in the
case of A78, a transitional [WO]-PG~1159 star, and in a similar object
with winds of even higher ionization, NGC~2371.  Through a comparison
of our models with the far-UV spectrum of the PG~1159-type CSPN K~1-16, we
have also demonstrated that the broad wind feature seen at this
wavelength in some PG~1159 objects originates not from \CIII\ (as
indicated in \citealp{werner:04}), but from \NeVII\ as well.

Our grid of models show that \NeVII\ \singlet 973.33 is a very strong
wind feature detectable at solar abundance levels, in contrast to
photospheric optical neon features \citep{werner:04}, in CSPN of high
stellar temperatures ($\Teff \gtrsim 85$~kK).  For the parameters of
NGC~2371 ($\log{\Mdot} = -7.1$~\Msunyr) and A78 ($\log{\Mdot} =
-7.3$~\Msunyr), the strength of the feature peaks for $130 \lesssim
\Teff \lesssim 145$~kK, and weakens dramatically for $\Teff \gtrsim
160$ (these cutoffs depend on the value of \Mdot\ and the neon
abundance).  For an enhanced neon abundance ($\XNe = 10\Xsun$), the
feature remains strong even in models of very high temperatures
($\Teff \gtrsim 165$~kK) or very low mass-loss rates ($\Mdot \simeq
1\E{-8}$~\Msunyr), while the lower \Teff\ limits remains approximately
the same.  We note here that the far-UV spectra of these
objects show the \NeVII\ feature being weaker that the \OVI\ line,
while in some PG~1159 stars (\eg, K~1-16 and Longmore~4), they are of
comparable strength.  Since PG~1159 stars represent a more advanced
evolutionary stage when the star is getting hotter and the wind is
fading, \NeVII\ \singlet 973 may be the last wind feature to disappear
if the atmosphere is enriched in neon.

In hydrogen-deficient objects, an enhanced neon abundance lends
credence to evolutionary models which have the star experiencing a
late helium shell flash, and predict a neon enrichment of about 20
times the solar value.  When saturated (\eg, in the case in NGC~2371),
the feature is insensitive to \Mdot, and only weakly sensitive to the
neon abundance.  Although models of these objects with enriched neon
abundances do result in better fits for our transition objects, the
sensitivity of the feature to $\Teff$ prevents us making a
quantitative statement regarding abundances based on this feature
alone.  Other far-UV/UV lines from \NeVI\ (at 2042, 2056, 2213, and
2229~\AA) and \NeVII\ (at 3644~\AA) which only appear in Ne-enriched
models, could in principle be used for this purpose, but we lack
observations in this range of sufficient quality/resolution to make a
quantitative assessment.  In the case of K~1-16, \NeVII\ \singlet 973
is unsaturated, and our models require an enhanced neon abundance to
fit it simultaneously with the \OVI\ \doublet 1032,38 profile.  This
result is in line with those of \citet{werner:94}, who derived a neon
abundance of 20 times the solar value for this object from analysis of
the \NeVII\ \singlet 3644 line.  The neon overabundance is further
evidence that this PG~1159 object has experienced the ``born-again''
scenario.

\NeVII\ \singlet 973 has diagnostic applications not only to late
post-AGB objects, but also for evolved massive stars. Evolutionary
models predict the surface neon abundance to vary dramatically as
Wolf-Rayet stars evolve (see, \eg, \citealp{meynet:05}).  For cooler
WR stars (such as those of the WN-type), the neon abundance can be
estimated from low-ionization features in the infrared and ultraviolet
(\eg, [\NeII ] 12.8$\mu$m, [\NeIII ] 15.5 $\mu$m, \NeIII\ \singlet
2553).  The \NeVII\ \singlet 973 feature may provide a strong neon
diagnostic for hotter, more evolved WR stars.  For example, the WO
star Sanduleak~2 has $\Tstar \simeq 150$~kK \citep{crowther:00}, and
appears to have a feature at that wavelength in a FUSE archive
spectrum.  Neon enhancements produced in massive stars may explain the
discrepancy in the $\:^{22}$Ne/$\:^{20}$Ne ratio between the solar
system and Galactic cosmic ray sources (see, \eg, \citealp{meynet:01}).

\acknowledgements

We are grateful to the anonymous referee for a careful reading of
the manuscript and their constructive comments.  We are indebted to the
members of the Opacity Project and Iron Project and to Bob Kurucz for
their continuing efforts to compute accurate atomic data, without
which this project would not have been feasible.  The SIMBAD database
was used for literature searches.  This work has been funded by NASA
grants NAG 5-9219 (NRA-99-01-LTSA-029) and NAG-13679.  The BEFS and
IUE data were obtained from the Multimission Archive (MAST) at the
Space Telescope Science Institute (STScI). STScI is operated by the
Association of Universities for Research in Astronomy, Inc., under
NASA contract NAS5-26555.

%\pagebreak
%-------references------------------------------------------------------

\clearpage
%---------------------figures-------------------------------------------

\clearpage

%---------------------tables--------------------------------------------

\input{tab1}

\input{tab2}

\input{tab3}

\end{document}

%% file: tab1.tex
\begin{table}[htbp]
\caption{Utilized Spectra}\label{tab:obs}
\begin{tabular}{ccccccc}
\hline
Star & Instrument & Data- & Date & Resolution & Aperture & Range \\
     &      & set    &      &        & (\arcsec)&  (\AA)  \\
\hline
NGC~2371 & FUSE & P1330301 & 02/26/00 & $\sim 0.05$~\AA & 30$\times$30 & 915--1180 \\
         & IUE & LWR04210 & 04/07/79  & 5--6~\AA & 10$\times$20 & 1975--3345 \\
Abell~78 & BEFS & BEFS2190 & 12/01/96 & $\sim 0.33$~\AA& 20 & 915-1222  \\
K 1-16 & FUSE & M1141201002 & 07/24/00 & $\sim 0.05$~\AA & 30$\times$30 & 915--1180 \\
\hline
\end{tabular}
\end{table}

%% file: tab2.tex
\begin{table}[htbp]
\caption{Superlevels and Levels for Model Ions}\label{tab:ion_tab}
\footnotesize
\begin{tabular}{lccccccccccc}
\hline
Element & I & II & III & IV & V & VI & VII & VIII & IX & X & XI\\
\hline
H  & 20/30 & 1/1 & & & & & & & & \\
He & 40/45 & 22/30 & 1/1\\
C  & & & 30/54 & 13/18 & 1/1\\
N  & & & & 29/53 & 13/21 & 1/1\\
O  & & & & 29/48 & 41/78 & 13/19 & 1/1\\
Ne & & & & 45/355 & 37/166 & 36/202 & 38/182 & 24/47 & 1/1 & \\
Si & & & & 22/33 & 1/1\\
P  & & & & 36/178 & 16/62 & 1/1\\
S  & & & & 51/142 & 31/98 & 28/58 & 1/1\\
Fe & & & & 51/294 & 47/191 & 44/433 & 41/254 & 53/324 & 52/490 &43/210 &1/1\\
\hline
\end{tabular}
\end{table}

%% file: tab3.tex
\begin{table}[htbp]
\caption{Stellar Parameters and Adopted Distances}\label{tab:mod_param_dist}
\begin{tabular}{cccccccc}
\hline
Star      & \Teff & \vinf & $D$ & \Rstar & $\log{L}$ &
$\log{\Mdot}$  \\
          & (kK)   & (\kms) & (kpc) & (\Rsun) & (\Lsun) & (\Msunyr) \\
\hline
NGC~2371$^a$  & $\textbf{135}^{+10}_{-15}$ & $\textbf{3700}\pm200$ & $1.5^b$ &
\textbf{0.09} &$\textbf{3.45}^{+0.12}_{-0.20}$ & $\textbf{-7.11}\pm0.30$ \\
Abell~78$^a$  & $\textbf{113}\pm8$ & $\textbf{3200}\pm50$ & $1.6^b$ &
\textbf{0.19} &$\textbf{3.73}^{+0.10}_{-0.13}$ & $\textbf{-7.33}^{+0.36}_{-0.13}$ \\
K~1-16  & \textbf{135} & \textbf{3700} & \textbf{2.05} & \textbf{0.11} & \textbf{3.6} & \textbf{-8.3} \\
\hline
\end{tabular}
\\
\begin{minipage}{\textwidth}
\scriptsize
\begin{trivlist}
\item (a): Stellar parameters from HB04
\item (b): Distance from \citet{cahn:92}
\item Parameters in \textbf{bold} are those adopted for the models in
  this paper.
\end{trivlist}
\end{minipage}
\end{table}

%% file: ms.bbl
\begin{thebibliography}{33}
\expandafter\ifx\csname natexlab\endcsname\relax\def\natexlab#1{#1}\fi

\bibitem[{Acker {et~al.}(1992)Acker, Marcout, Ochsenbein, Stenholm, \&
  Tylenda}]{acker:92} Acker, A., Marcout, J., Ochsenbein, F.,
  Stenholm, B., \& Tylenda, R. 1992, Strasbourg-ESO Catalogue of
  Galactic Planetary nebulae (ESO Publication)

\bibitem[{Bianchi \& Grewing(1987)}]{bianchi:87b}
Bianchi, L. \& Grewing, M. 1987, \aap, 181, 85

\bibitem[{Cahn {et~al.}(1992)Cahn, Kaler, \& Stanghellini}]{cahn:92}
Cahn, J.~H., Kaler, J.~B., \& Stanghellini, L. 1992, \aaps, 94, 399

\bibitem[{Crowther(2000)}]{crowther:00}
Crowther, P.~A. 2000, \aap, 356, 191

\bibitem[{Crowther {et~al.}(1998)Crowther, De~Marco, \& Barlow}]{crowther:98}
Crowther, P.~A., De~Marco, O., \& Barlow, M.~J. 1998, \mnras, 296, 367

\bibitem[{Crowther {et~al.}(2002)Crowther, Dessart, Hillier, Abbott, \&
  Fullerton}]{crowther:02}
Crowther, P.~A., Dessart, L., Hillier, D.~J., Abbott, J.~B., \& Fullerton,
  A.~W. 2002, \aap, 392, 653

\bibitem[{De~Marco \& Soker(2002)}]{demarco:02}
De~Marco, O. \& Soker, N. 2002, \pasp, 114, 602

\bibitem[{Grevesse \& Sauval(1998)}]{grevesse:98}
Grevesse, N. \& Sauval, A.~J. 1998, \ssr, 85, 161

\bibitem[{Grewing \& Neri(1990)}]{grewing:90}
Grewing, M. \& Neri, R. 1990, \aap, 236, 223

\bibitem[{Herald \& Bianchi(2002)}]{herald:02}
Herald, J.~E. \& Bianchi, L. 2002, \apj, 580, 434

\bibitem[{Herald \& Bianchi(2004{\natexlab{a}})}]{herald:04a}
---. 2004{\natexlab{a}}, \apj, 611, 294

\bibitem[{Herald \& Bianchi(2004{\natexlab{b}})}]{herald:04b}
---. 2004{\natexlab{b}}, \apj, 609, 378

\bibitem[{Herwig(2001)}]{herwig:01}
Herwig, F. 2001, \apss, 275, 15

\bibitem[{Herwig {et~al.}(1999)Herwig, Bl\"{o}cker, Langer, \&
  Driebe}]{herwig:99}
Herwig, F., Bl\"{o}cker, T., Langer, N., \& Driebe, T. 1999, \aap, 349, 5

\bibitem[{Hillier {et~al.}(2003)Hillier, Lanz, Heap, Hubeny, Smith, Evans,
  Lennon, \& Bouret}]{hillier:03}
Hillier, D.~J., Lanz, T., Heap, S.~R., Hubeny, I., Smith, L.~J., Evans, C.~J.,
  Lennon, D.~J., \& Bouret, J.~C. 2003, \apj, 588, 1039

\bibitem[{Hillier \& Miller(1998)}]{hillier:98}
Hillier, D.~J. \& Miller, D.~L. 1998, \apj, 496, 407

\bibitem[{Hillier \& Miller(1999)}]{hillier:99b}
---. 1999, \apj, 519, 354

\bibitem[{Iben \& McDonald(1995)}]{iben:95}
Iben, I. \& McDonald, J. 1995, in LNP, Vol. 443, White Dwarfs, ed. D.~Koester
  \& K.~Werner (Heidelberg: Springer), 48

\bibitem[{Kaler {et~al.}(1993)Kaler, Stanghellini, \& Shaw}]{kaler:93}
Kaler, J.~B., Stanghellini, L., \& Shaw, R.~A. 1993, \aap, 279, 529

\bibitem[{Koesterke \& Werner(1998)}]{koesterke:98b}
Koesterke, L. \& Werner, K. 1998, \apj, 500, 55

\bibitem[{Kruk \& Werner(1998)}]{kruk:98}
Kruk, J.~W. \& Werner, K. 1998, \apj, 502, 858

\bibitem[{Lindeberg(1972)}]{lindeberg:72}
Lindeberg, S. 1972, Uppsala Univ. Inst. Phys. Report UUIP-759, 1

\bibitem[{Luo \& Pradhan(1989)}]{luo:89a}
Luo, D. \& Pradhan, A.~K. 1989, \prb, 22, 3377

\bibitem[{Maciel(1984)}]{maciel:84}
Maciel, W.~J. 1984, \aaps, 55, 253

\bibitem[{{Meynet} {et~al.}(2001){Meynet}, {Arnould}, {Paulus}, \&
  {Maeder}}]{meynet:01}
{Meynet}, G., {Arnould}, M., {Paulus}, G., \& {Maeder}, A. 2001, Space Science
  Reviews, 99, 73

\bibitem[{{Meynet} \& {Maeder}(2005)}]{meynet:05}
{Meynet}, G. \& {Maeder}, A. 2005, \aap, 429, 581

\bibitem[{Morgan {et~al.}(2003)Morgan, Parker, \& Cohen}]{morgan:03}
Morgan, D.~H., Parker, Q.~A., \& Cohen, M. 2003, \mnras, 346, 719

\bibitem[{{Opacity Project Team}(1995)}]{opacity:95}
{Opacity Project Team}. 1995, The Opacity Project, Vol.~1 (Bristol: Institute
  of Physics Publications)

\bibitem[{{Opacity Project Team}(1997)}]{opacity:97}
---. 1997, The Opacity Project, Vol.~2 (Bristol: Institute of Physics
  Publications)

\bibitem[{Peach {et~al.}(1988)Peach, Saraph, \& Seaton}]{peach:88}
Peach, G., Saraph, H.~E., \& Seaton, M.~J. 1988, \prb, 21, 3669

\bibitem[{Pe\~{n}a {et~al.}(2004)Pe\~{n}a, Hamann, Ruiz, Peimbert, \&
  Peimbert}]{pena:04}
Pe\~{n}a, M., Hamann, W.-R., Ruiz, M.~T., Peimbert, A., \& Peimbert, M. 2004,
  \aap, 419, 583

\bibitem[{Seaton(1987)}]{seaton:87}
Seaton, M.~J. 1987, \prb, 20, 6363

\bibitem[{Tully {et~al.}(1990)Tully, Seaton, \& Berrington}]{tully:90}
Tully, J.~A., Seaton, M.~J., \& Berrington, K.~A. 1990, \prb, 23, 3811

\bibitem[{Werner {et~al.}(2003)Werner, Dreizler, Koesterke, \&
  Kruk}]{werner:03}
Werner, K., Dreizler, S., Koesterke, L., \& Kruk, J.~W. 2003, in IAU Symp.,
  Vol. 209, Planetary Nebulae: Their Evolution and Role in the Universe, ed.
  M.~Dopita \& S.~Kwok (San Fransisco: ASP), 239

\bibitem[{Werner \& Rauch(1994)}]{werner:94}
Werner, K. \& Rauch, T. 1994, \aap, 284, 5

\bibitem[{Werner {et~al.}(2004)Werner, Rauch, Reiff, Kruk, \&
  Napiwotzki}]{werner:04}
Werner, K., Rauch, T., Reiff, E., Kruk, J.~W., \& Napiwotzki, R. 2004, \aap,
  427, 685

\bibitem[{Young {et~al.}(2003)Young, Del~Zanna, Landi, {et~al.}}]{young:03}
Young, P.~R., Del~Zanna, G., Landi, E., {et~al.} 2003, \apjs, 144, 135


\end{thebibliography}
